\documentclass[prl,twocolumn,nofootinbib]{revtex4}
\usepackage{graphicx, epsfig}
\usepackage{color}
\usepackage{mathrsfs}
\usepackage{bm}
\usepackage{amsmath,amssymb}
\usepackage[caption=false]{subfig}
\usepackage{hyperref}
\usepackage[normalem]{ulem}
\usepackage{mathtools}

\newcommand{\be}{\begin{equation}}
\newcommand{\ee}{\end{equation}}
\newcommand{\ba}{\begin{eqnarray}}
\newcommand{\ea}{\end{eqnarray}}

\newcommand{\Lqcd}{\Lambda_{\rm QCD}}
\newcommand{\tqcd}{t_{\rm QCD}}
\newcommand{\fpq}{f_{\rm PQ}}
\newcommand{\fpqunit}{f_*}

\newcommand{\tbreak}{t_{\rm f}}
\newcommand{\pbh}{p_{\rm bh}}

\begin{document}
\title{Lunar Mass Black Holes from QCD Axion Cosmology}
\author{Tanmay Vachaspati$^{\dag *}$}
\affiliation{
$^\dag$Physics Department, Arizona State University, Tempe, AZ 85287, USA. \\
$^*$Maryland Center for Fundamental Physics, University of Maryland,
                    College Park, MD 20742, USA.
}

\begin{abstract}
\noindent
In the QCD axion scenario, a network of domain walls bounded by cosmic strings fragments into
pieces. As these fragments collapse, some of them will form black holes. With standard
QCD axion parameters, the black holes will have lunar masses 
($M_{\rm bh} \sim 10^{-8}\, {\rm M}_\odot$). Even though their number density is difficult to 
estimate, arguments suggest that they can constitute a reasonable fraction of the critical 
cosmological density.
\end{abstract}

\maketitle

{\it Note:} More recent simulations~\cite{Fleury:2015aca} show that the string-wall network will fragment
earlier than what is estimated below. Then the wall pieces will be smaller and the chance
of forming a black hole will be lower. Thus the QCD axion scenario is unlikely to
lead to an interesting number of black holes. However other axion models, with different parameters
and cosmological evolution, might lead to black holes in larger numbers and the physics discussed in 
this paper could still be of interest.

\

The axion was proposed as a way to understand the smallness of the neutron
electric dipole moment even when theoretical considerations would suggest
a large value. (For a review see~\cite{Kim:2008hd}). An unexpected benefit
is that axions can play the role of cold dark matter in cosmology~\cite{Marsh:2015xka}.
At the same time, an unintended consequence of the QCD axion model is the existence 
of a network of cosmic strings in the early universe, at temperatures above the QCD 
scale. At the QCD scale additional domain walls are created that connect the strings in
the network. In the usual picture the domain walls shrink and cause the 
network to collapse and annihilate into axion radiation, thus leaving no consequential 
signature of their one time existence~\cite{Vilenkin:1982ks,Kibble:1982dd}. In this paper 
we point out that the collapse of the string network can also produce lunar mass
black holes. However, their number density is difficult to estimate. 
The arguments are similar to those 
already given for black hole formation from collapsing cosmic string 
loops~\cite{Hawking:1990tx,Fort:1993zb} and also from domain
walls produced during inflation~\cite{Deng:2016vzb,Rubin:2001yw}\footnote{The black 
holes we discuss are unrelated to axion stars and axion 
miniclusters~\cite{Tkachev:1986tr,Kolb:1993zz,Ballesteros:2016euj} that have recently 
been constrained by microlensing searches~\cite{Fairbairn:2017dmf}.}.

The timeline of the QCD axion starts at the so-called Peccei-Quinn energy scale,
$f_{\rm PQ}$, when a global U(1) symmetry breaks spontaneously to the
trivial group. The phase of the Peccei-Quinn complex scalar field is a Goldstone boson 
and is called the axion. The U(1) symmetry breaking generates a network of cosmic 
global strings (reviewed in~\cite{Vilenkin:2000jqa}). 
The tension of the global strings, $\mu$, is given by
\be
\mu \approx \pi \, \ln(\fpq L) \fpq^2
\label{stringtension}
\ee
where $L$ is the typical inter-string separation and is bounded by the cosmic
horizon 
and $\fpq$ is taken to be the vacuum expectation value of the 
Peccei-Quinn complex
scalar field. The logarithmic factor is weakly dependent on the length $L$.
For example, if $L$ is of order the horizon at the QCD temperature,
$\tqcd \approx 10^{-3}~{\rm s}$, and $\fpq \approx 10^{11}~{\rm GeV}$,
we have $\ln(\fpq L) \approx 70$. 

The Peccei-Quinn scale, $\fpq$, is constrained by 
cosmology~\cite{Hagmann:2000ja,2010PhRvD..82l3508W,Hiramatsu:2012gg,Visinelli:2014twa,
Klaer:2017ond} to be between $10^{10}-10^{12}~{\rm GeV}$ and we will take
\be
\fpq = 3\times 10^{10}\, {\rm GeV} \, f_* 
\ee
with $f_* \approx 1$ a free parameter.

The string network evolves under the forces of string tension, Hubble expansion, 
and backreaction from Goldstone boson radiation; the network also reconfigures 
itself due to intercommutations when strings intersect. The Goldstone boson radiation 
from global strings is quite efficient~\cite{Vilenkin:1986ku} and we expect the long 
strings to not be very curved. The typical coherence scale of the strings will be 
assumed to be on the cosmic horizon scale.

\begin{figure}
      \includegraphics[width=0.42\textwidth,angle=0]{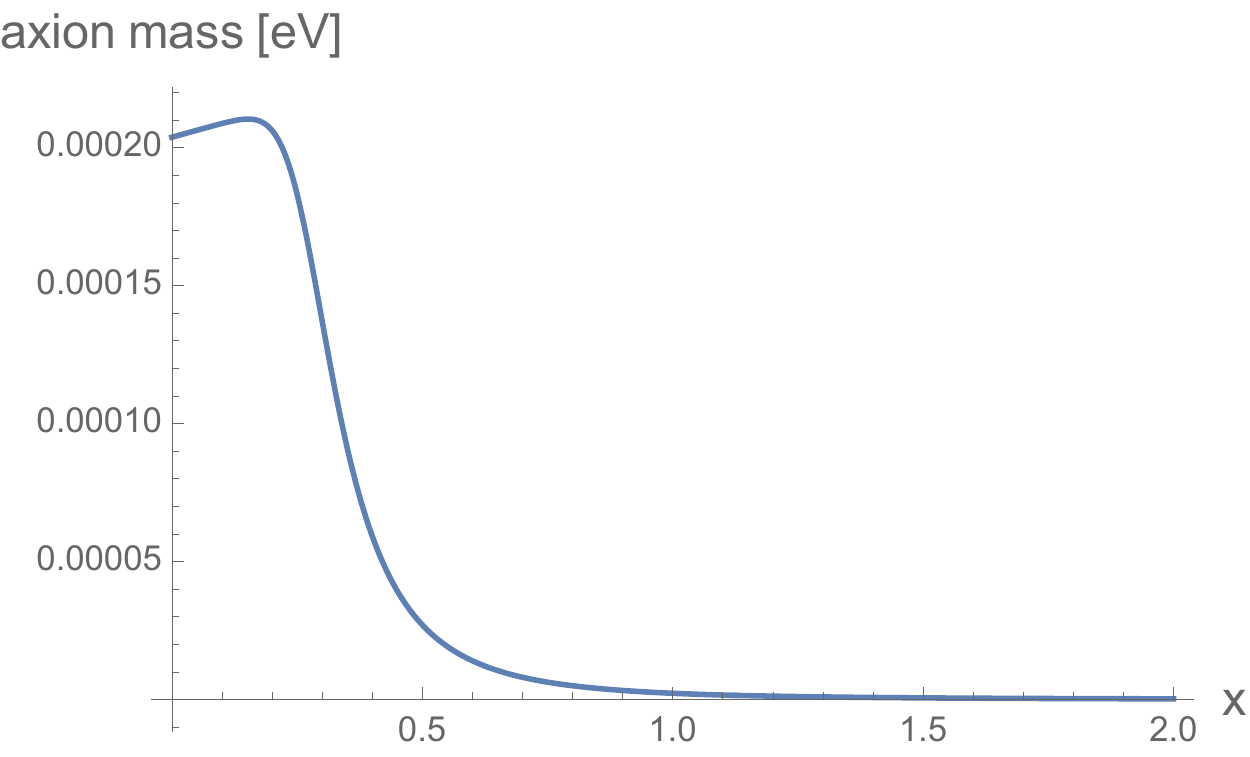}
  \caption{Mass of axion in eV versus $x=T/\Lambda$ for 
  $\fpq=3\times 10^{10}~{\rm GeV}$~\cite{2010NuPhB.829..110W}.}
\label{maplot}
\end{figure}

The string network evolves freely until  the cosmic temperature drops to 
about the QCD energy scale, $\Lqcd \approx 200\, {\rm MeV}$ when the
axions rapidly acquire a mass~\cite{2010NuPhB.829..110W,Borsanyi:2016ksw}. 
Ref.~\cite{2010NuPhB.829..110W} gives a convenient formula for the
axion mass as a function of cosmic temperature,
\be
m_a^2 (T) \frac{\fpq^2}{\Lambda^4} = 
\begin{cases}
1.46 \times 10^{-3} 
      \frac{1+0.5 x}{1+(3.53\, x)^{7.48}}, & x < 1.125
\\ 
\\
1.68\times 10^{-7} x^{-6.68}, & x \ge 1.125
\end{cases}
\label{ma2T}  
\ee
where  $x\equiv T/\Lambda$, 
$\Lambda=400~{\rm MeV} \equiv 2\Lqcd$. 
In Fig.~\ref{maplot} we plot the temperature dependence of the axion mass.

The temperature dependence causes the axion mass to turn on in time at about 
the QCD cosmological epoch.
To determine the time-dependence we use the cosmic time-temperature relation,
\be
\frac{1}{4t^2} = H^2 = \frac{8\pi G}{3} \frac{\pi^2}{30} g_* T^4
\ee
where $g_* \approx 70$ counts the relativistic degrees of freedom. This gives,
\be
\frac{1}{t} = 27.8 \frac{T^2}{m_P}
\label{tT}
\ee
where the Planck mass $m_P=1.22\times 10^{19}~{\rm GeV}$.

With a non-zero mass of the axion, the phase of the complex scalar field, namely
the axion denoted by the field $a$, develops a potential that is approximately described 
by the sine-Gordon potential. The Lagrangian is
\be
L = \fpq^2 \left [ \frac{1}{2} (\partial_\mu a)^2 - m_a^2 (1-\cos(a)) \right ]
\ee
The potential is periodic under $a \to a+2\pi n$ where $n$ is any
integer. Thus there are domain wall solutions that interpolate between the 
minima of the potential. 
The wall tension (energy per unit area) in the sine-Gordon model is
\be
\sigma = 8 m_a f_{\rm PQ}^2.
\label{sigma}
\ee
A precision analysis shows slight departures of the true axion potential
from the sine-Gordon model~\cite{diCortona:2015ldu} and that the correct 
wall tension is $8.97\, m_a f_{\rm PQ}^2$. We will disregard this small 
difference and continue to describe axion dynamics by the sine-Gordon
model.

There are three relevant epochs for the evolution. First is the time, denoted
$t_H$, when the tension in the axion walls starts to dominate over the Hubble expansion.
This is given by $m_a (t_H) t_H =1$ and we find that the temperature
at this epoch is $T_H \simeq 2~{\rm GeV}$. This is also the time at which 
the force on the network due to domain walls becomes more important than the 
tension in horizon-size strings {\it i.e.} $\sigma > \mu / t$.
A second relevant time is when the string-wall 
network fragments into isolated pieces. We will denote this time $\tbreak$ and
$\tbreak > t_H$ because the tension in the walls has to overcome Hubble expansion
for the network to fragment. A third relevant time, denoted $t_a$, is when the axion 
has acquired its asymptotic mass.
From the plot of $m_a(T)$ in Fig.~\ref{maplot} we see that this happens at a temperature 
$T_a \approx 0.2 \Lambda$. The numerical values for $T_a$ and the corresponding epoch are
\be
T_a \approx 0.2 \Lambda \approx 80~{\rm MeV}, \ \ t_a \approx 4.5\times 10^{-5}~{\rm s}
\label{Tata}
\ee
and the axion mass from this time on is,
\be
m_{a,0} \approx 2.0\times 10^{-4}~{\rm eV}\ f_*^{-1}
\label{ma0}
\ee


The next task will be to estimate $\tbreak$. 
Early simulations of the string-wall network performed in Ref.~\cite{Hiramatsu:2012gg}
indicate that the network does not fragment until after $t_a$ (see Figs.~2 and 4 of 
Ref.~\cite{Hiramatsu:2012gg}). However, the more recent analysis of 
Ref.~\cite{Klaer:2017ond} shows earlier fragmentation at a time $\sim 0.1 t_a$. {check}

The estimates in Eqs.~(\ref{Tata}) and (\ref{ma0}) also give
\be
m_{a,0} t_a \sim 1.4 \times 10^7 \, f_*^{-1} 
\label{ma0ta}
\ee
which says that the size $\sim t_a$ of the walls is much larger than their width 
$\sim m_a^{-1}$. 

Once the network has fragmented, the pieces, that we call ``membranes'', start to collapse 
due to tension and may collapse into black holes. However, a membrane will also lose energy 
into radiation as it collapses. Then there is competition between the rate of collapse and
the rate of radiation. This process has been considered for local cosmic 
strings~\cite{Hawking:1990tx} and for global cosmic string 
loops~\cite{Fort:1993zb}. For circular gauge cosmic strings, the dominant emission is to 
gravitational radiation. This is quite weak and leads to a significant range of parameters that 
can give black holes. For circular global cosmic string loops, Goldstone boson radiation
is very efficient and the parameter space for black hole formation is very restricted.
However, since the axion has a mass, our membranes are not like global strings
and black hole formation requires a separate study. We will first give a rough estimate
for when black holes can form and then analyze the collapse of spherical sine-Gordon
walls in more detail (similar to the analysis in Ref.~\cite{Widrow:1989vj} for $Z_2$ walls).
 
Consider a membrane in the shape of a circular disk that starts contracting from an
initial radius $R_0$ that is close to the horizon size $t_a$. 
To form a black hole, the membrane must collapse so that its radius at some later time
$t_{\rm bh}$ satisfies $R(t_{\rm bh})=2GM$ where $M=\sigma \pi R_0^2$ and $\sigma$ 
is the wall tension in Eq.~(\ref{sigma}) with $m_a$ replaced by $m_{a,0}$. However, there 
is also a second constraint: the Schwarzschild radius $R(t_{\rm bh})$ must be larger than 
the width of the wall $\sim m_{a,0}^{-1}$, otherwise radiation will become important once
$R(t) < m_{a,0}^{-1}$ as discussed in~\cite{Fort:1993zb}.
Therefore the condition for black hole formation is
\be
2 G \sigma \pi R_0^2 >  R(t_{\rm bh}) \gtrsim m_{a,0}^{-1}.
\label{bhcondition0}
\ee
This condition is to be taken as a rough guide. For example, the $m_{a,0}^{-1}$ on the 
right-hand side ignores the Lorentz contraction of the wall and string as the system collapses.
For a disk membrane, this will make the string thinner by the inverse of the boost factor
but the wall thickness will not change; for a spherical wall, the wall will get Lorentz
contracted.
We continue with Eq.~(\ref{bhcondition0}) for now, as it may be more relevant to
the case of a circular disk, but will do a more careful numerical analysis for a collapsing 
spherical wall below.
The black hole formation condition can also be written as
\be
\frac{R_0}{t_a} \gtrsim \frac{1}{\sqrt{18\pi}} \frac{m_P}{m_{a,0} t_a \fpq} \simeq 4
\label{bhcondition}
\ee
This estimate shows that we do not expect black hole formation  in general. However, the estimate is 
close enough that we expect black holes to form with some reduced probability.
For example, geometric factors for the shape of the wall can contribute to this condition -- we 
could have considered a membrane with a larger surface area than that of a disk. Also, the typical 
scale of the network may be set by the Hubble scale instead of the cosmic time which would make
the initial size somewhat bigger than $t_a$. And the membranes might at first be
stretched due to Hubble flow which would also make them bigger.

\begin{figure}
    \includegraphics[width=0.23\textwidth,angle=0]{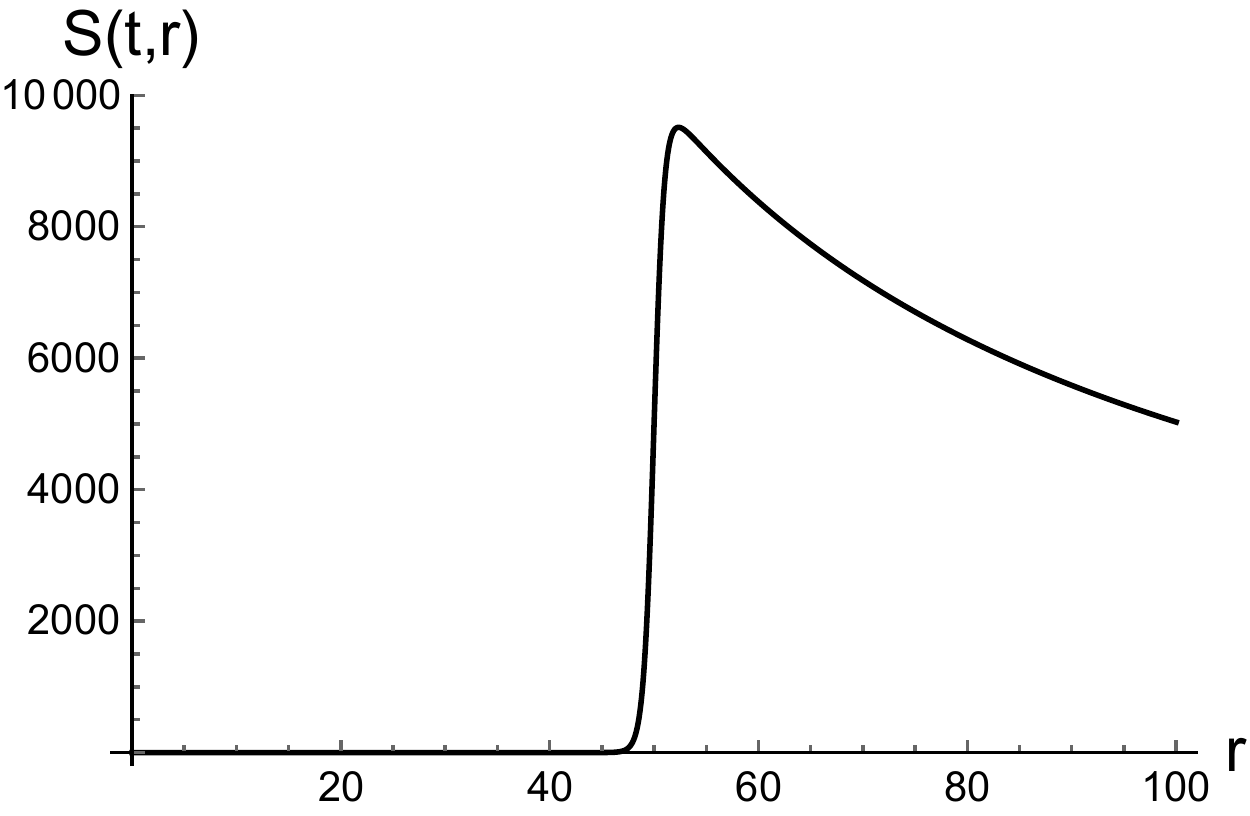}
     \includegraphics[width=0.23\textwidth,angle=0]{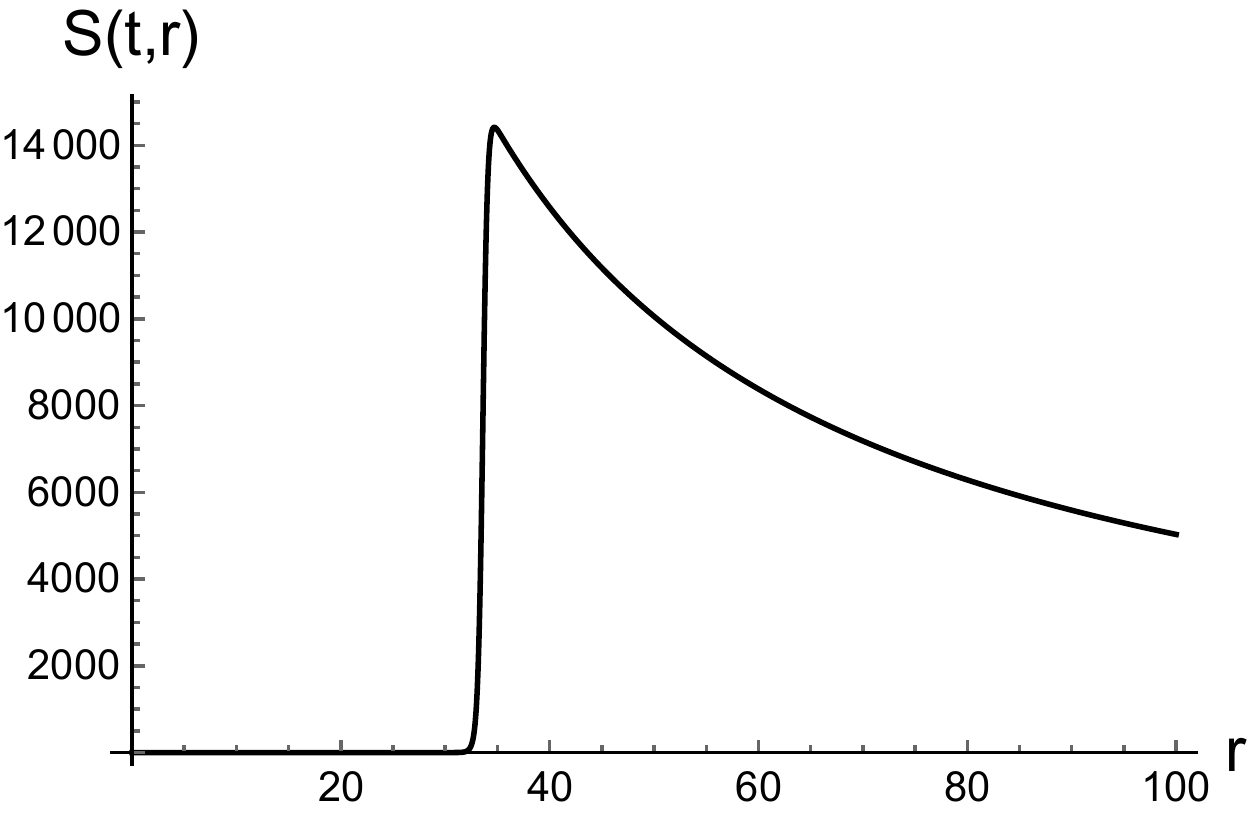}
      \includegraphics[width=0.23\textwidth,angle=0]{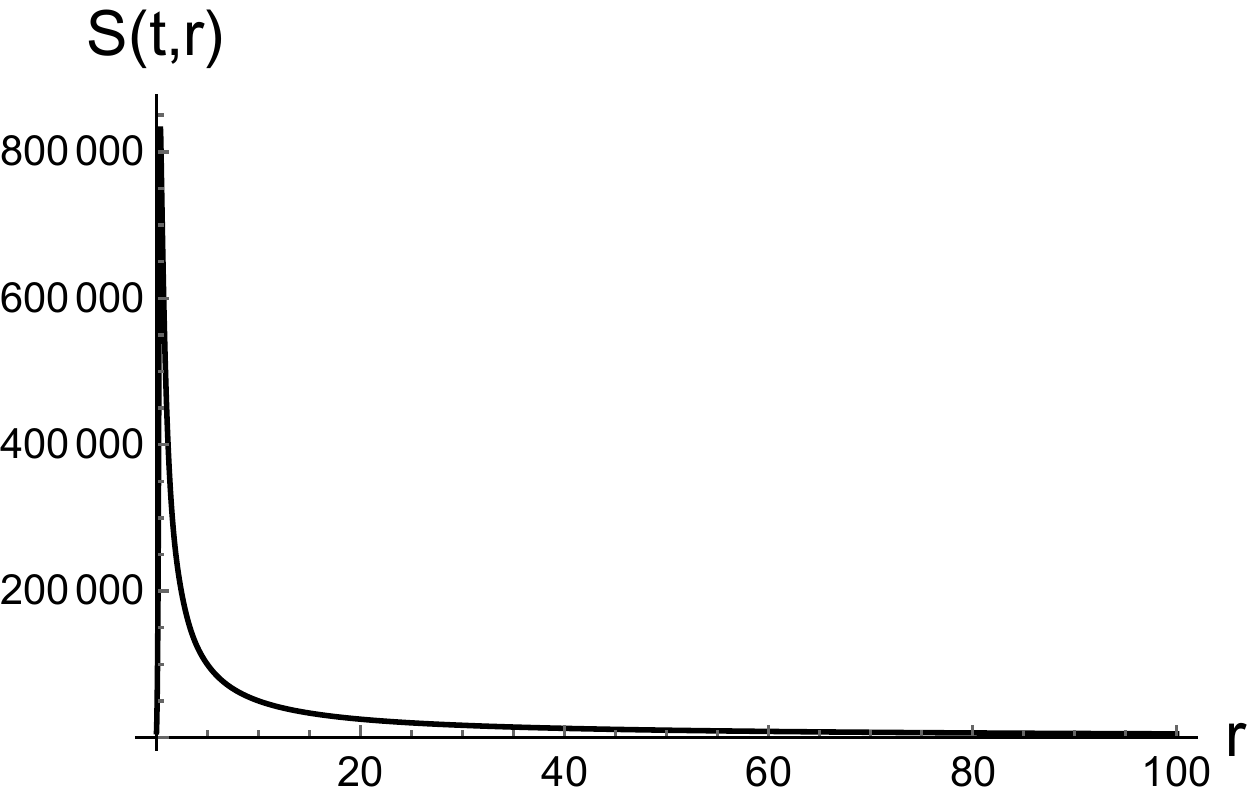}
       \includegraphics[width=0.23\textwidth,angle=0]{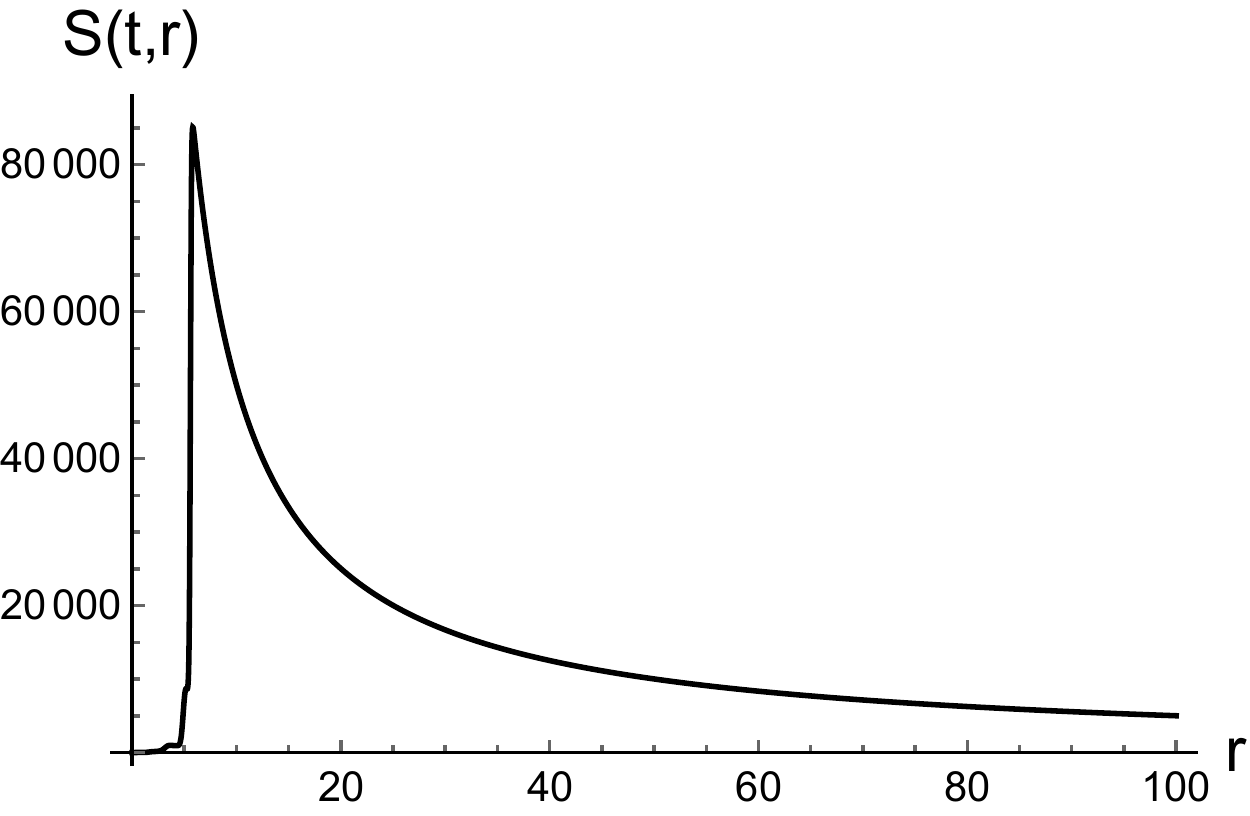}
  \caption{Plots of the surface gravity function, $S(t,r)$, at four different
  times. Top left plot is close to the initial time, top right at an intermediate time,
  bottom left at a time very close to when $S(t,r)$ attains its maximum
  value, and bottom right after the maximum has been attained. Note
  that the maximum value of $S(t,r)$ in the bottom left plot is 
  $\sim 800,000$, while the
  peak values in the other three plots are at $\sim 8000$, $14000$ and $80,000$.
 }
 \label{SvstAndr}
\end{figure}

\begin{figure}
\begin{center}
    \includegraphics[width=0.45\textwidth,angle=0]{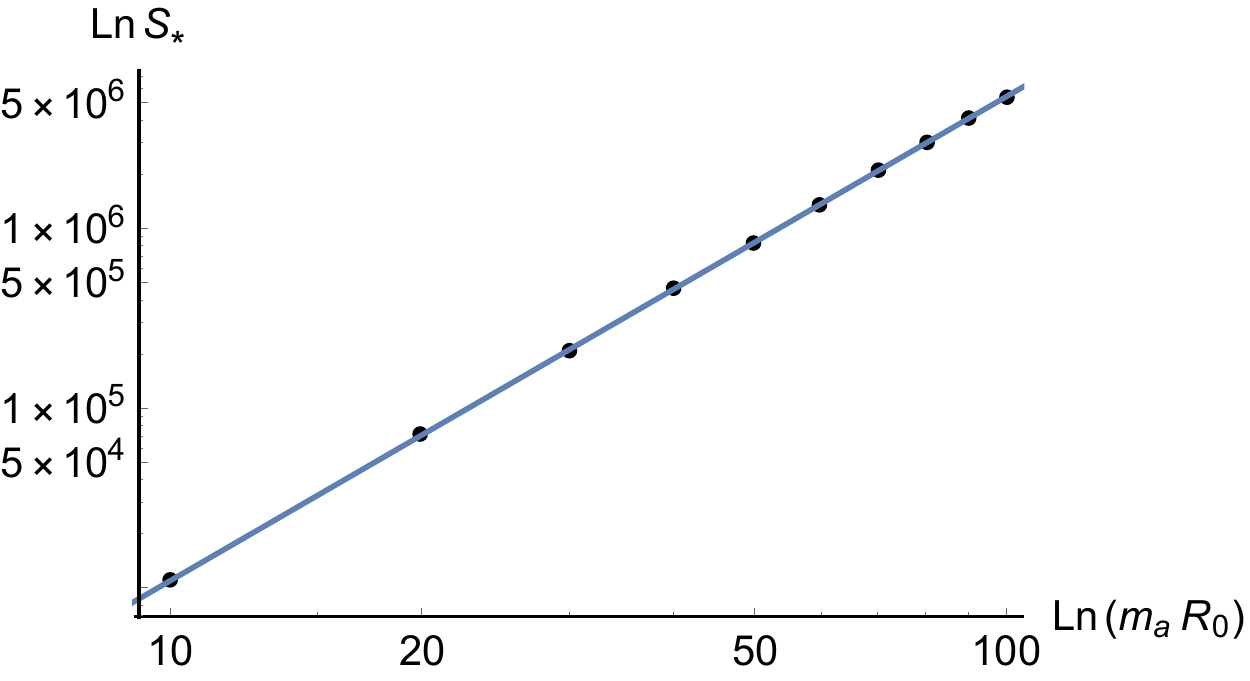}
\end{center}
  \caption{Log-log plot (dots) of the numerically evaluated maximum surface gravity of the collapsing 
  spherical wall versus initial radius of the wall. The straight line fit shown is
 $\ln(S_*) = 3.08805 + 2.69554 \ln (m_a R_0)$.
 }
 \label{SvsR0}
\end{figure}

The above estimate gives us a rough idea for when black holes will form. We can examine
the particular case of collapse of a spherical domain wall in much greater detail by numerically
solving its equation of motion,
\be
\partial_t^2 a = \nabla^2 a - \sin(a)
\ee
where we have set $m_{a,0}=1$ by rescaling coordinates.
The initial field for the spherical domain wall can be written as~\cite{Vachaspati:2006zz}
\be
a(t=0,r) = 4  [ \tan^{-1}(e^{r+R_0}) + \tan^{-1}(e^{-r+R_0}) ] - 2\pi
\ee
where $r$ is the (rescaled) spherical radial coordinate. We also 
start with a static spherical wall so ${\dot a}(t=0,r)=0$.
We calculate the energy $E(r)$ contained within a sphere of radius $r$ and then
evaluate a rescaled surface gravity at time $t$ and radius $r$, 
$S(t,r)=2E(t,r)/(f_{\rm PQ}^2 r)$. (The factor of $f_{\rm PQ}^2$ is included
to cancel out this same factor in $E(t,r)$ so that $S$ does not depend on 
$\fpq$.) At any given time, $S(t,r)$ 
increases at small $r$ and decreases as $1/r$ at very large $r$ as
seen in Fig.~\ref{SvstAndr}. Therefore $S(t,r)$ has a
maximum, $S(t,r_*)$, at a certain radius, $r_*$. As the wall collapses, $S(t,r_*)$
increases at first as the wall becomes more compact, but eventually it decreases
due to wall annihilation and radiation. So $S(t,r_*)$ has a maximum that we
denote by $S_*$ at some time $t_*$,
\be
S_* \equiv {\rm max}_{(t,r)} \left ( \frac{2E(t,r)}{f_{\rm PQ}^2 r} \right )
\ee
This is the maximum surface gravity attained during the collapse of the wall and
only depends on the initial radius, $R_0$, of the wall. 
The power law dependence of $S_*$ on $R_0$ is shown in Fig.~\ref{SvsR0} and 
gives 
\be
S_*=21.9~ (m_{a,0} R_0)^{2.7}.
\ee
This formula is independent of $f_{\rm PQ}$. A black hole will form if $2GE/r > 1$
which is equivalent to
\be
S_*=21.9~ (m_{a,0} R_0)^{2.7}  > \frac{m_P^2}{f_{\rm PQ}^2}
\ee
Therefore to form a black hole we need to start with a spherical domain wall of radius
\be
R_0 > m_{a,0}^{-1} 
            \left ( \frac{m_P^2}{21.9 \, f_{\rm PQ}^2} \right )^{1/2.7}
\approx 7.6\times 10^5 \, m_{a,0}^{-1} \, f_*^{-0.74}
\label{Rc}
\ee
Comparison with the estimate of the horizon size in Eq.~(\ref{ma0ta}) shows that 
the critical radius for black hole formation from spherical walls is $\sim 0.1 t_a$, 
instead of $\sim 4t_a$ based on the estimate of Eq.~(\ref{bhcondition}). 
Thus, depending on their shape, large but still sub-horizon walls 
can collapse to form black holes.

The typical black hole mass at formation is given by the energy in a horizon size
membrane. Using Eq.~(\ref{ma0ta}) we get,
\be
M(t_a) = \sigma \pi t_a^2 \approx 2 \times 10^{-8}~{\rm M}_\odot ~ \fpqunit 
\label{bhmass}
\ee
where $1\, {\rm M}_\odot \approx 2\times 10^{33}\, {\rm gms}$. For comparison, the mass
of the Earth is $\approx 3\times 10^{-6}~{\rm M}_\odot$ and the mass of the Moon is
$\approx 4\times 10^{-8}~{\rm M}_\odot$.

The growth of primordial black holes has been of long-standing interest and has
recently been discussed in Refs.~\cite{Guedens:2002sd,Deng:2016vzb}. A basic
picture of the growth is given by
\be
{\dot M} = \rho A
\label{dotM}
\ee
where $\rho$ is the ambient radiation energy density and $A$ is the area of
the black hole: $A= 4\pi R^2 = 16\pi G^2M^2$. The differential equation 
(\ref{dotM}) can be
solved and the growth of the black hole is determined by the 
ratio of its mass to the energy within the horizon at $t_a$,
\be
\frac{M(t_a)}{M_H(t_a)} \sim 10^{-8}.
\ee
As this fraction is very small, the growth is negligible and can be ignored, giving
\be
M(t_0) \approx 2 \times 10^{-8}\, {\rm M}_\odot ~ \fpqunit .
\label{bhmasst0}
\ee
The Schwarzschild radius of one of these black holes is 
$R_S \approx 6\times 10^{-8}~ {\rm km} ~\fpqunit \sim 0.1~{\rm mm} ~\fpqunit$.

The number density of black holes depends on how many membranes undergo
gravitational collapse. Not every membrane will be sufficiently symmetric, and
angular momentum can prevent the membrane from contracting to its Schwarzschild
radius. There is also a chance that a convoluted collapsing membrane will fragment further,
however this requires a self-intersection along an entire closed curve. (A generic
self-intersection will occur at two points and that will change the topology of the wall 
without leading to fragmentation.) 
Let us denote by $\pbh$ the probability that a large membrane, for which radiative 
losses can be ignored, collapses to a black hole. So $\pbh$ absorbs our ignorance 
of the membrane angular momentum and fragmentation probability.

In terms of $\pbh$ the mass density in black holes at time $t_a$ is
\be
\rho_{\rm bh} (t_a) \sim \frac{\pbh \, M(t_a)}{4\pi t_a^3/3}
\ee
and their energy density relative to the critical density, $\rho_c = 3/(32\pi G t^2)$, at 
formation is
\be
\Omega_{\rm bh} (t_a) = \frac{\rho_{\rm bh} (t_a)}{\rho_c(t_a)} 
\approx 2 \times 10^{-8} \pbh ~\fpqunit .
\ee
The relative energy density grows with scale factor in the radiation era and at the
present epoch is
\be
\Omega_{\rm bh} (t_0) = \Omega_{\rm bh} (t_a)  \left ( \frac{T_a}{T_{\rm eq}} \right ) 
\approx 2 \, \pbh \, \fpqunit
\label{Omegat0}
\ee
where $T_{\rm eq}\approx 1~{\rm eV}$ is the temperature at the epoch of matter-radiation
equality. 

Estimates have been made for black hole formation probability from cosmic 
string loops~\cite{Hawking:1987bn,Polnarev:1988dh}. 
A general argument proposed by Rees
(reviewed in \cite{Vilenkin:2000jqa}) is based
on the angular momentum barrier to gravitational collapse -- a string loop can only
collapse to a black hole of mass $M$ if its angular momentum is less than the 
maximum allowed for a black hole, $J_{\rm max} = GM^2$. In the case of global strings,
we expect the strings to be less curved than local strings, and the membranes to be
relatively flat, as also seen in simulations (see Fig.~2 of~\cite{Hiramatsu:2012gg}). 
We will assume
that a membrane inherits all its angular momentum from the motion of strings that
intersect. Since the strings move at relativistic velocities and have size $R_0$,
the angular momentum of a membrane is $J \sim \mu R_0^2$. 
\be
\frac{J_{\rm max}}{J} \sim \frac{GM^2}{\mu R_0^2} \sim 4\times 10^{-3}.
\ee
First following Rees, we assume that every component of the angular momentum is 
independent and uniformly distributed, and we require that all components be
smaller than $J_{\rm max}$. Then we estimate
\be
\pbh \sim  \left ( \frac{J_{\rm max}}{J} \right )^3 \sim 10^{-7}
\ee
which gives $\Omega_{\rm bh} (t_0) \sim 10^{-7} \fpqunit$.

On the other hand, we do not expect all three components of the angular
momentum to be independent. The velocity of a string has to be perpendicular
to the tangent direction to the string. So if a membrane is formed from the
intersection of two relatively straight strings, we expect that the angular
momentum vector component along the strings will be large but the components
in the orthogonal directions will be small. In this case a more suitable upper
bound is
\be
\pbh \sim  \frac{J_{\rm max}}{J}  \sim 2 \times 10^{-3}
\ee
which gives $\Omega_{\rm bh} (t_0) \sim 4\times 10^{-3}\, \fpqunit$.

Clearly these are tentative estimates of $\pbh$ and need to be investigated more carefully.
However, it is likely that $\Omega_{\rm bh}(t_0)$ is much smaller than 1 and will not violate 
microlensing constraints which give $\Omega_{\rm bh} (t_0) \lesssim 0.01$ 
(see 
Fig.~20 of \cite{Niikura:2017zjd}). If our estimates of 
$\pbh$ are too conservative and $\Omega_{\rm bh} (t_0) \sim 0.01$ then these black holes 
may be a significant component of the cosmic dark matter~\cite{2016PhRvL.116t1301B} in 
addition to the usual coherent axionic dark matter. 

The mass spectrum of black holes will be determined by the mass distribution of 
membranes once the string-wall network fragments. Drawing an analogy with the 
better studied monopole-string systems in
which the length distribution of strings is exponentially suppressed \cite{Vilenkin:2000jqa}, 
we expect that the mass spectrum of membranes will be exponentially suppressed by the
initial area of the membrane. Then the resulting black hole mass spectrum will also
be exponentially suppressed by the mass of the black hole and only the 
lowest mass black holes will be relevant.
Further, the recent analysis in Ref.~\cite{2016PhRvL.116t1301B} for the black hole
merger rate within galaxy halos will apply to axion black holes as well. The analysis
assumes a dark matter density for the black holes but the resulting merger rate is 
independent of the black hole mass.

With the parameters of the QCD axion the black hole masses are too small by a factor of 
$\sim 10^9$ to be the black holes seen by LIGO. If we consider an ``axion-like particle'' 
(ALP) instead of the QCD axion, and if the physics of the ALP also leads to a string-wall 
network that fragments, the resulting black holes could have significantly higher masses.
It would be worth examining black hole formation in a concrete ALP model.


\acknowledgements
I thank Kohei Kamada, David Marsh, Rashmish Mishra,  Raman Sundrum and Alex Vilenkin 
for comments and discussions, and especially Shmuel Nussinov for explaining the constraints
arising from Bondi accretion.
This work is supported by the U.S. Department of Energy, 
Office of High Energy Physics, under Award No. DE-SC0013605 at Arizona State University.

\bibstyle{aps}
\bibliography{axionbh}

\end{document}